\documentclass[conference]{IEEEtran}
\IEEEoverridecommandlockouts
\usepackage[nolist]{acronym}
\usepackage{tabularx}
\usepackage{booktabs}
\usepackage{todonotes}
\begin{acronym}[AoI]
\acro{AoI}{Area of Interest}
\end{acronym}
\begin{acronym}[ARQ]
\acro{ARQ}{Automatic Repeat Request}
\end{acronym}
\begin{acronym}[CACC]
\acro{CACC}{Cooperative Adaptive Cruise Control}
\end{acronym}
\begin{acronym}[CAM]
\acro{CAM}{Cooperative Awareness Message}
\end{acronym}
\begin{acronym}[CBR]
\acro{CBR}{Channel Busy Ratio}
\end{acronym}
\begin{acronym}[CA]
\acro{CA}{Cooperative Awareness}
\end{acronym}
\begin{acronym}[CBF]
\acro{CBF}{Contention-Based Forwarding}
\end{acronym}
\begin{acronym}[CCAM]
\acro{CCAM}{Cooperative, Connected and Automated Mobility}
\end{acronym}
\begin{acronym}[C-ITS]
\acro{C-ITS}{Cooperative Intelligent Transport Systems}
\end{acronym}
\begin{acronym}[CP]
\acro{CP}{Collective Perception}
\end{acronym}
\begin{acronym}[CPM]
\acro{CPM}{Collective Perception Message}
\end{acronym}
\begin{acronym}[DCC]
\acro{DCC}{Decentralized Congestion Control}
\end{acronym}
\begin{acronym}[DEN]
\acro{DEN}{Decentralized Environmental Notification}
\end{acronym}
\begin{acronym}[DENM]
\acro{DENM}{Decentralized Environmental Notification Message}
\end{acronym}
\begin{acronym}[DPD]
\acro{DPD}{Duplicate Packet Detection}
\end{acronym}
\begin{acronym}[DPL]
\acro{DPL}{Duplicate Packet List}
\end{acronym}
\begin{acronym}[EDCA]
\acro{EDCA}{Enhanced Distributed Channel Access}
\end{acronym}
\begin{acronym}[ETSI]
\acro{ETSI}{European Telecommunications Standards Institute}
\end{acronym}
\begin{acronym}[FCD]
\acro{FCD}{Floating Car Data}
\end{acronym}
\begin{acronym}[ITS]
\acro{ITS}{Intelligent Transport Systems}
\end{acronym}
\begin{acronym}[LocT]
\acro{LocT}{Location Table}
\end{acronym}
\begin{acronym}[LocTE]
\acro{LocTE}{Location Table entry}
\end{acronym}
\begin{acronym}[MC]
\acro{MC}{Maneuver Coordination}
\end{acronym}
\begin{acronym}[MCO]
\acro{MCO}{Multi-channel Operation}
\end{acronym}
\begin{acronym}[MCM]
\acro{MCM}{Maneuver Coordination Message}
\end{acronym}
\begin{acronym}[mmWave]
\acro{mmWave}{millimiter wave}
\end{acronym}
\begin{acronym}[PAM]
\acro{PAM}{Platooning Awareness Message}
\end{acronym}
\begin{acronym}[PCM]
\acro{PCM}{Platooning Control Message}
\end{acronym}
\begin{acronym}[PDR]
\acro{PDR}{Packet-delivery Ratio}
\end{acronym}
\begin{acronym}[RadCom]
\acro{RadCom}{Radar-Based Communication}
\end{acronym}
\begin{acronym}[RSU]
\acro{RSU}{Road-side Unit}
\end{acronym}
\begin{acronym}[RHW]
\acro{RHW}{Road Hazard Warning}
\end{acronym}
\begin{acronym}[SHB]
\acro{SHB}{Single-Hop Broadcasting}
\end{acronym}
\begin{acronym}[S-CBR]
\acro{S-CBR}{Smoothed Channel Busy Ratio}
\end{acronym}
\begin{acronym}[TC]
\acro{TC}{Traffic Class}
\end{acronym}
\begin{acronym}[TDC]
\acro{TDC}{Transmit Data Rate Control}
\end{acronym}
\begin{acronym}[TLS]
\acro{TLS}{Transport Layer Security}
\end{acronym}
\begin{acronym}[TPC]
\acro{TPC}{Transmit Power Control}
\end{acronym}
\begin{acronym}[TRC]
\acro{TRC}{Transmit Rate Control}
\end{acronym}
\begin{acronym}[TTL]
\acro{TTL}{Time-to-Live}
\end{acronym}
\begin{acronym}[VA]
\acro{VA}{Vulnerable Road User awareness}
\end{acronym}
\begin{acronym}[VANET]
\acro{VANET}{Vehicular ad hoc Network}
\end{acronym}
\begin{acronym}[V2X]
\acro{V2X}{Vehicle-to-anything}
\end{acronym}
\begin{acronym}[V2V]
\acro{V2V}{Vehicle-to-Vehicle}
\end{acronym}
\usepackage{cite}
\usepackage{amsmath,amssymb,amsfonts}
\usepackage{algorithmic}
\usepackage{graphicx}
\usepackage{textcomp}
\usepackage{xcolor}
\def\BibTeX{{\rm B\kern-.05em{\sc i\kern-.025em b}\kern-.08em
    T\kern-.1667em\lower.7ex\hbox{E}\kern-.125emX}}
\begin{document}

\title{The Smart Highway to Babel: the Coexistence of Different Generations of Intelligent Transport Systems
}

\author{\IEEEauthorblockN{Oscar Amador}
\IEEEauthorblockA{\textit{School of Information Technology} \\
\textit{Halmstad University}\\
Halmstad, Sweden \\
email: oscar.molina@hh.se}\\
\IEEEauthorblockN{Mar\'ia Calder\'on}
\IEEEauthorblockA{\textit{Departamento de Ingeniería de Sistemas Telemáticos} \\ \textit{ETSI Telecomunicación} \\
\textit{Universidad Politécnica de Madrid}\\
Madrid, Spain \\
email: maria.calderon@upm.es}\\
\and
\IEEEauthorblockN{Ignacio Soto}
\IEEEauthorblockA{\textit{Departamento de Ingeniería de Sistemas Telemáticos} \\ \textit{ETSI Telecomunicación} \\
\textit{Universidad Politécnica de Madrid}\\
Madrid, Spain \\
email: ignacio.soto@upm.es}\\
\IEEEauthorblockN{Manuel Urue\~na}
\IEEEauthorblockA{\textit{Escuela Superior de Ingenieros y Tecnolog\'{\i}a} \\
\textit{Universidad Internacional de la Rioja}\\
Logro\~no, Spain \\
email: manuel.uruena@unir.net}\\
}

\maketitle

\begin{abstract}
The gap between technology readiness level in \ac{C-ITS} and its adoption and deployment has caused a phenomenon where at least two types of network access technologies have to coexist. Furthermore, for the case of the \ac{ETSI} Intelligent Transport Systems protocols, work is being completed in Release 2 of the specification while Release 1 deployments are still underway. This, coupled with industry and consumer trends in the vehicle industry, is bound to cause a scenario where fully C-ITS-enabled vehicles have to coexist with \textit{non}-C-ITS road users and, at the very least, with different versions of C-ITS. In this paper, we analyze the performance in terms of efficiency and safety of two releases of the ETSI GeoNetworking protocol and we discuss possible paths to tackle the upcoming compatibility and coexistence problems. 
\end{abstract}


\section{Introduction}
The use of \ac{C-ITS} to maximize road safety and traffic efficiency has been one of the cornerstones upon which future mobility is built. The final stage of \ac{CCAM} depends on the presence of \ac{C-ITS} on all roads and at all times, exchanging information and coordinating their maneuvers~\cite{ecVisionZero}.

The road to \ac{CCAM} is divided in three different fronts: \textit{connection} (the ability to exchange information through networks), \textit{cooperation} (the protocols that define how intelligent vehicles react to information and each other's actions), and \textit{automation} (the level of human intervention on the driving task). These fronts have particular stages (e.g., levels of automation~\cite{SAELevels}), but they share common stages, such as the \textit{Days} in Vision Zero~\cite{ecVisionZero}. These Days (1--4) are incremental steps toward the realization of full \ac{CCAM}:

\begin{itemize}
    \item On Day 1, \textit{awareness} starts, and vehicles share their status using messages like \ac{CAM} and \ac{DENM} (i.e., in the framework established by the \ac{ETSI});
    \item on Day 2, \textit{cooperation} starts, and vehicles exchange information from their sensors using, e.g., \acp{CPM};
    \item on Day 3, road users communicate their intentions; and
    \item on Day 4, road users execute coordinated maneuvers.
\end{itemize}

These days take into account the evolution of technology. For example, in the \textit{connection} front, Day 1 considers the use of \acp{VANET} supported on cellular communications (i.e., LTE) or in WiFi (e.g., ETSI~ITS-G5, based on IEEE~802.11p). From Day 2 onward, C-ITSs expect the use of evolved technologies (e.g., 5G, 802.11bd, and technologies beyond these two). The choice between cellular or WiFi is the first hurdle towards the harmonic coexistence of different types of intelligent vehicles, and \ac{ETSI} develops media-dependent protocols for both approaches~\cite{etsiMediaDependentG5,etsiMediaDependentLTE}. Thus, manufacturers and transportation authorities are given the chance to select one or many technologies.

However, industry an consumer patterns are likely to cause a scenario where vehicles that are produced in 2023, with the technological features present this year, will share the road with fully CCAM-enabled vehicles in 2050~\cite{Netherlands2050}. Even now, figures from the industry show that the average age for a vehicle in Europe ranges from 12 to 14.7 years for cars and trucks respectively, and some countries have even larger mean values~\cite{fleetage}. This means that is highly likely to have a fleet with 1) different technological capabilities, and 2) different versions of the same technology.

In this paper, we present the effect of the coexistence of two versions of one safety-critical protocol: Release 1 of ETSI~\ac{CBF}~\cite{etsiNewGeoNetworking}, and the changes proposed to Release 2, which were originally presented in~\cite{Amador2022} and~\cite{S-FoT+:2023}. We evaluate efficiency metrics such as the number of transmissions and its variation with larger penetration rates of the newer protocol in scenarios where a message has to be distributed within a Destination Area. Finally, we discuss the likely scenarios for coexistence and possible compatibility between two versions of one protocol.

The rest of the paper is organized as follows: in Section~\ref{sec:background}, we present the two releases of the \ac{ETSI} \ac{CBF} protocol; in Section~\ref{sec:experiment}, we perform an experimental assessment of the penetration rate of the updated protocol on effectivity and efficiency; Section~\ref{sec:discussion} presents a discussion on scenarios and alternatives to paliate the problem of having a mixed fleet; and finally, conclusions and future work are presented in Section~\ref{sec:conclusion}.

\section{Background}
\label{sec:background}

\subsection{ETSI ITS Architecture}

\begin{figure}[tb]
    \centering
    \includegraphics[width=\linewidth]{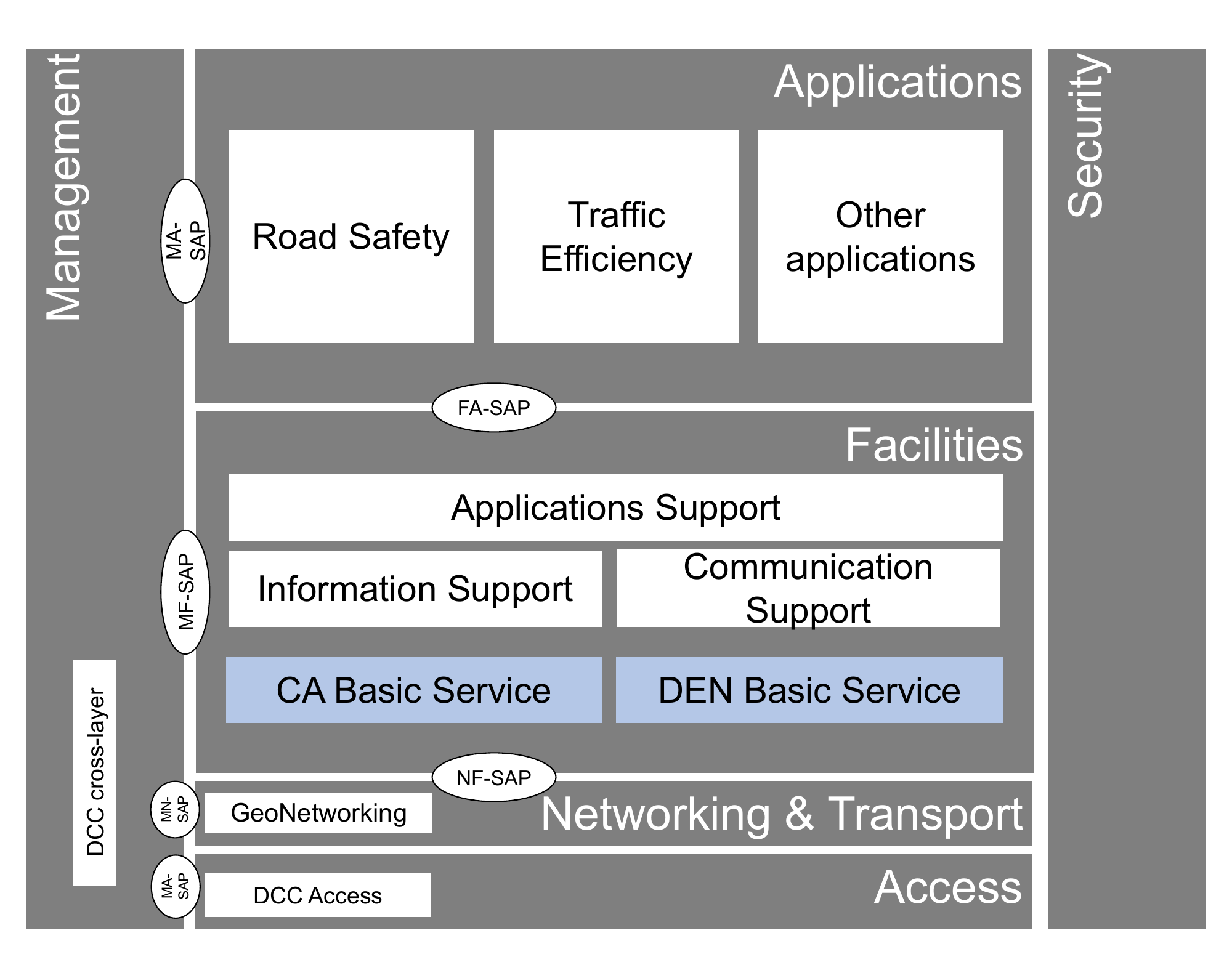}
    \caption{ETSI ITS Architecture.}
    \label{fig:architecture}
\end{figure}

Figure~\ref{fig:architecture} shows the layers and entities of the \ac{ETSI} ITS architecture. At the very top, the Application layer hosts systems that pursue the goals of all \acp{C-ITS} --- road safety and traffic efficiency --- as well as other functions (e.g., related to infotainment). These applications are supported by the Facilities layer, e.g., by safety-critical Day 1 services like the \ac{CA} and \ac{DEN} basic services. These services exchange messages with other nodes (vehicles and the infrastructure) that allow applications fulfill their roles: for example, a \ac{DENM} warns road users about roadworks ahead of the road, and an application can suggest or take a new route. 

Messages are generated by services at the Facilities layer and then get sent down the stack to the Networking \& Transport layer. Depending on the use case and requirements from applications, a message can be broadcast to neighbors one hop away (i.e., \ac{SHB}), or towards a specific area of interest (Destination Area). The latter is achieved through GeoNetworking~\cite{etsiNewGeoNetworking}. In either case, packets are encapsulated and sent down to the Access layer for transmission.

The Access layer executes Medium Access Control as well as Congestion Control functions. This layer accommodates both WiFi-based and cellular-based access technologies. For the case of WiFi-based access (i.e., \ac{ETSI} ITS-G5), channel occupation (i.e., \ac{CBR}) is measured at this layer and, using \ac{DCC}~\cite{etsiNewDcc}, each station calculates the share of the medium it can use, which ranges from 0.06\% to 3\% of the medium, or a message rate between 1 and 40\,Hz. This means that, even in extremely low congestion conditions, consecutive messages must wait in the \ac{DCC} queues for at least 25\,ms between each dequeuing. From these queues, frames are then sent to the \ac{EDCA} queues where they wait for their time to contend for access to the medium.

The road a message takes from generation to transmission and the possible bottleneck or sinkhole effects that different phenomena, e.g., at the Access layer, can have on protocol performance is accounted for by \ac{ETSI} protocols. E.g., a \ac{CAM} can only be generated if the message rate is less or equal to the one allowed by \ac{DCC}. However, the appearance of new services and the expected effect of having a high number of nodes in proximity of each other has prompted the research community to study these effects continuously~\cite{amador2020decentralized}.

\subsection{GeoNetworking in ETSI ITS}
\label{subsec:GeoNetworking}

Routing protocols in conventional computer networks rely on Layer 3 addresses to send data between hosts in remote locations. This is typically achieved through IP addressing. In the context of \acp{VANET}, where use cases sometimes require the dissemination of information to a given area, geographical awareness is required for a routing protocol. Hence, GeoNetworking functionalities are included, e.g., in the Networking \& Transport layer of the \ac{ETSI} ITS protocol stack~\cite{etsiNewGeoNetworking}. GeoNetworking allows for messages to reach a Destination Area without the need of maintaining a record of the network addresses of nodes in that area, which would be difficult due to the dynamic nature of vehicular networks.

ETSI defines mechanisms to broadcast information to a geographical Destination Area when:
\begin{itemize}
    \item the source is outside the Destination Area and the message has to arrive in it (e.g., using Greedy Forwarding or \ac{CBF}); or
    \item the message originates from or arrives into the Destination Area and is disseminated using \ac{CBF} or Simple Forwarding.
\end{itemize}

Non-Area mechanisms are out of the scope of this paper, but we can summarize Greedy Forwarding as a mechanism where each hop selects its farthest known neighbor and determines it as the next hop toward the Destination Area. These type of mechanisms have been widely studied, and the ETSI-defined version of Greedy Forwarding is evaluated in-depth in~\cite{riebl2021} and~\cite{Sandonis2016}.

Regarding Area forwarding mechanisms, Simple Forwarding can be described as a brute-force mechanism where every node that receives a message forwards it immediately (i.e., simple flooding). \ac{CBF}, on the other hand, makes receivers start a \textit{contention} timer that is proportional to their distance from the last hop before they decide to forward the message. If they listen to a forwarding while they are waiting for their timer to expire, they cancel the timer and drop their copy of the packet.

\subsubsection{Inefficiencies in Release 1 of ETSI CBF}
Efforts from the research community have evaluated the performance of \ac{ETSI} \ac{CBF}. While the theoretical frame which supports \ac{CBF} makes it more optimal than, e.g., simple forwarding, the way it interacts with other layers in the \ac{ETSI} ITS architecture causes phenomena that affect its efficiency.

The interaction between \ac{ETSI} \ac{CBF} and the \ac{DCC} mechanism at the Access layer causes an undesired effect: even if the \ac{CBF} timer expires, and the decision to forward the packet is made, if there is congestion in the channel or if another packet has just been transmitted, the forwarding is stopped at the \ac{DCC} queues (for \ac{ETSI} ITS-G5) or the scheduler (for C-V2X). This means that the actual transmission may not occur when \ac{CBF} has decided, and this phenomenon can occur in any station, so even if a copy of the message is received during contention, it is not guaranteed that it comes from an optimal forwarder. Furthermore, Release 1 of \ac{ETSI} GeoNetworking relies \ac{DPD} to \ac{CBF}, so, if a backlogged message from a DCC-affected forwarder is received at a neighbor which had already forwarded or even cancelled its copy will enter the loop once again.

\subsubsection{ETSI CBF Release 2}
The issues with \ac{DPD} and the effect of \ac{DCC} on Release 1 for \ac{ETSI}~\ac{CBF} had been studied widely in the literature~\cite{riebl2021,Kuhlmorgen2020,Paulin2015}. Yet, it was the work in~\cite{Amador2022} and~\cite{S-FoT+:2023} that was presented to \ac{ETSI} as a change request that was iterated and matured before it reached the necessary consensus to be Release 2 of \ac{ETSI}~\ac{CBF}.

The differences in Release 2 of Area \ac{CBF} are:
\begin{itemize}
    \item The inclusion of \ac{DPD} inside the \ac{CBF} algorithm.
    \item Interfacing with the cross-layer \ac{DCC} mechanism to offer awareness of the time before \ac{DCC} allows the next transmission, and account for it when calculating the contention timer (optional for cellular-based communications).
    \item A procedure to determine if a copy received during contention actually comes from a better forwarder.
    \item An updated timer formula to account for receptions beyond the maximum expected distance.
\end{itemize}

However, since Release 2 services might have different requirements and characteristics, it is not clear if Release 1 nodes will be able to receive messages originating from Release 2 nodes, even for safety-critical applications. If this is the case, and nodes executing Release 2 of \ac{ETSI} \ac{CBF} coexist with nodes executing Release 1, there might be effects on awareness and efficiency metrics. In the following section, we evaluate these effects in Area \ac{CBF} in a highway scenario.

\section{Experimental Evaluation of Coexistent Releases}
\label{sec:experiment}

\subsection{Simulation Scenario}
We evaluate the effect of different ratios of nodes executing Release 1 and 2 of \ac{ETSI}~\ac{CBF} in a highway scenario where a vehicle is stationary on the shoulder of a road. It starts sending \acp{DENM}~\cite{etsiDEN} at 1\,Hz with a Destination Area covering 4\,km of a road with 4 lanes per direction. The vehicular density is 30~veh/km on each lane. We take measurements for 30\,s after a warm-up period of 120\,s. We evaluate:

\begin{enumerate}
    \item \textbf{\ac{PDR}:} the number of successful individual receptions of a message in the Destination Area divided by the total number of vehicles in the area at the time of \ac{DENM} generation.
    \item \textbf{Number of transmissions:} how many transmissions (i.e., from the source and forwarders) have occurred. 
\end{enumerate}

Our toolkit consists of the OMNET++-based simulator Artery~\cite{Artery}, which implements the \ac{ETSI} ITS protocol stack using Vanetza and Veins~\cite{Veins} for the physical model of \ac{ETSI} ITS-G5. Mobility is controlled by SUMO~\cite{sumo2012}. A set of vehicles execute \ac{ETSI} \ac{CBF}Release 1~\cite{etsiNewGeoNetworking}, and an increasing number of vehicles (see the penetration rate parameter) execute the improvements included in Release 2 as described in~\cite{S-FoT+:2023}. In our set-up, and due to the nature of the message (i.e., \ac{RHW}), we consider Release 2 and Release 1 messages to be mutually understandable. The rest of the parameters are specified in Table~\ref{tbl:simpars}.

\begin{table}[h]
	\centering
  	\caption{Simulation Parameters}
	\begin{tabular}{| l | l |}
		\hline
		\textbf{Parameter}  & \textbf{Values} \\
		\hline
		Access Layer protocol & ITS-G5 (IEEE 802.11p) \\
		Channel bandwidth & 10\,MHz at 5.9\,GHz \\
		Data rate & 6\,Mbit/s \\
            DCC & ETSI Adaptive DCC \\
		Transmit power & 20\,mW \\
		Path loss model & Two-Ray interference model\textcolor{black}{~\cite{Sommer:2012}} \\
		Maximum transmission range& 1500\,m \\
		CAM packet size & 285 bytes \\
            \textcolor{black}{CAM generation frequency} & \textcolor{black}{1--10~Hz (ETSI CAM~\cite{etsiCA})} \\
		CAM Traffic Class & TC2 \\
		DENM packet size & 301 bytes \\
		DENM Traffic Class & TC0 (Source) and TC3 (Forwarders) \\
		DENM lifetime & 10\,s\\
		DPL size & 32 packet identifiers per Source\\
            Default Hop Limit & 10\\
            Rel. 2 penetration rate & 0, 25, 50, 75, 100\%\\
		\hline
	\end{tabular}
	\label{tbl:simpars}
\end{table}

\subsection{Results}

\begin{figure}[tbh]
    \centering
    \includegraphics[width=1\linewidth]{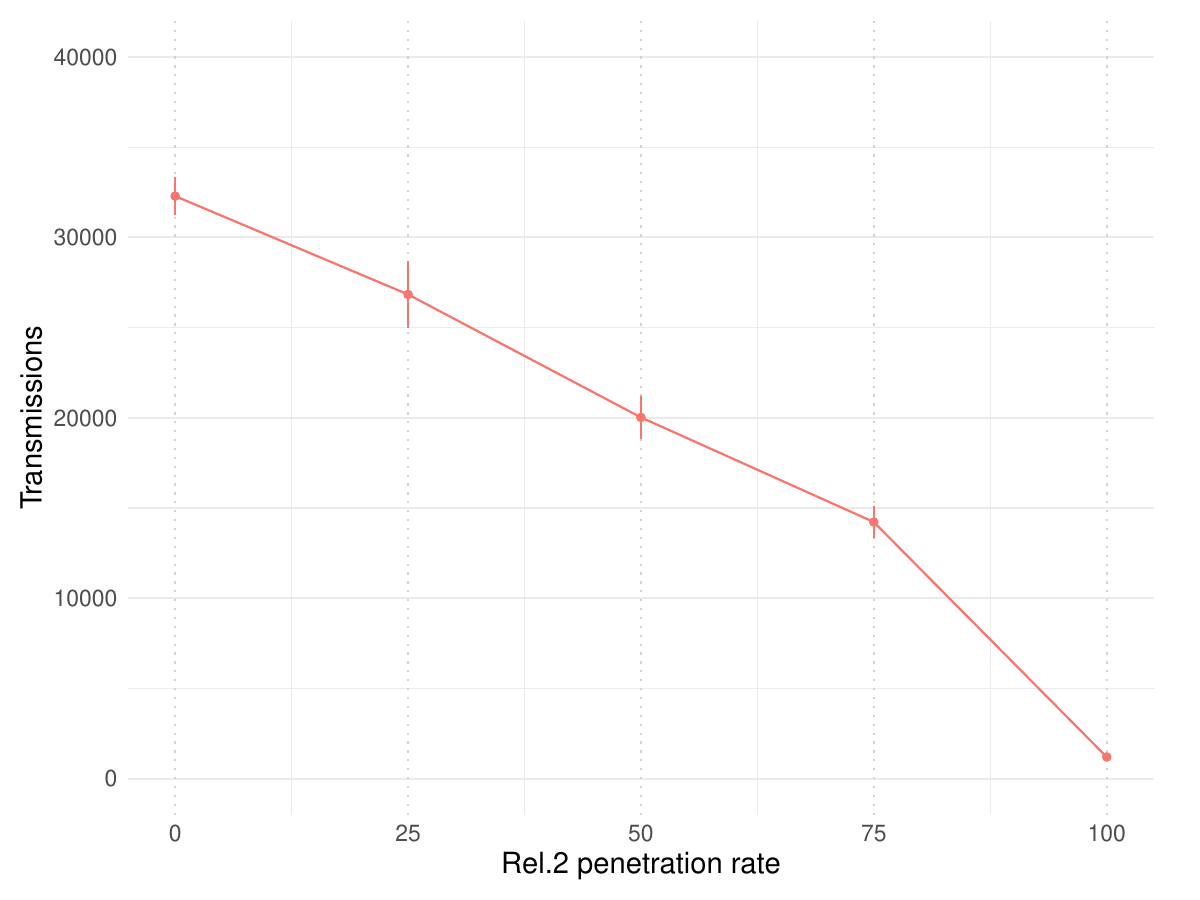}
    \caption{Number of transmissions in different Release 2 penetration rates.}
    \label{fig:txd}
\end{figure}

\begin{figure}[tb]
    \centering
    \includegraphics[width=\linewidth]{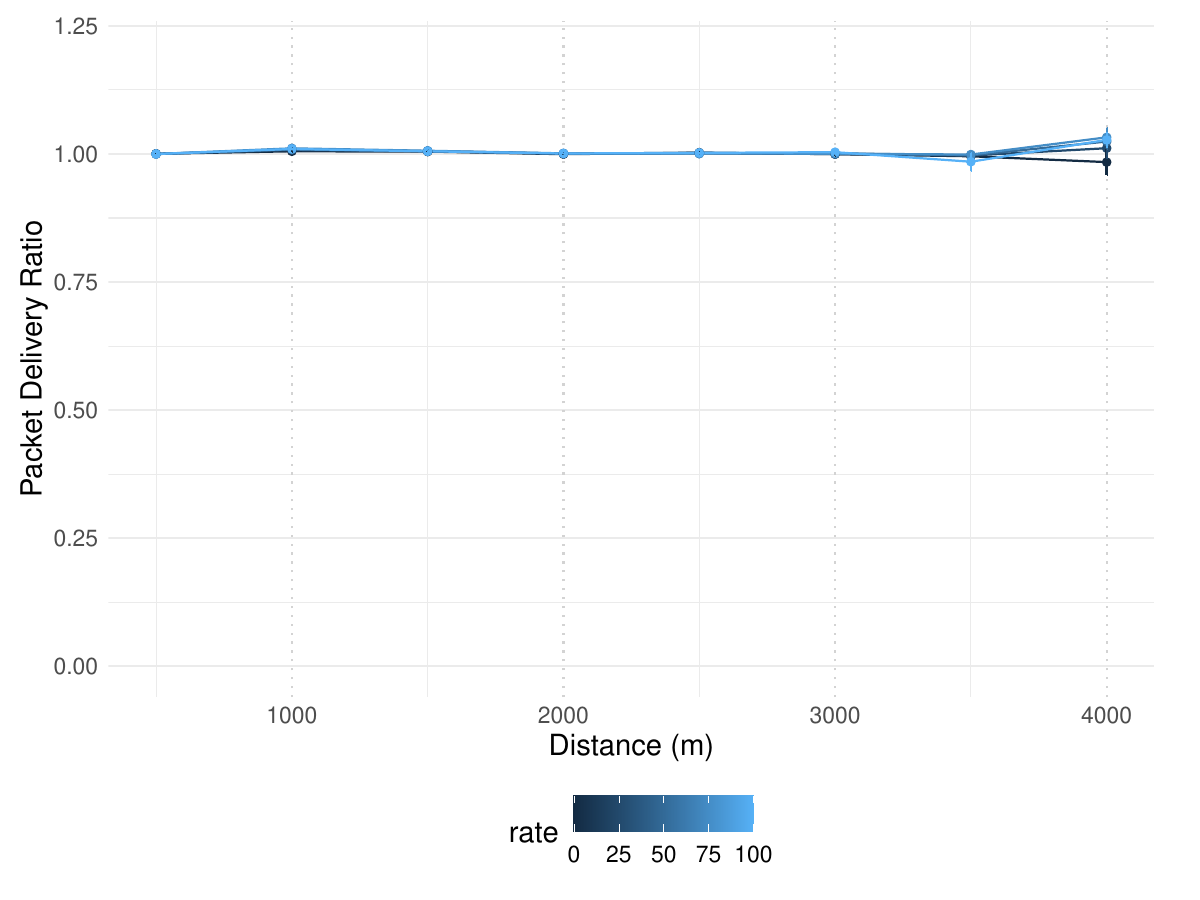}
    \caption{Packet Delivery Ratio for different Release 2 penetration rates.}
    \label{fig:pdrdist}
\end{figure}

Figure~\ref{fig:txd} shows the effect of even a minority portion of nodes executing a non-optimized protocol. There is beyond an order of magnitude in executed transmissions between the 0\% and the 25\% penetration rate for Release 2. From there, there is a linear increase until the almost 30:1 ratio between Release 1 and Release 2 in line with the results in~\cite{Amador2022} and~\cite{S-FoT+:2023}. 

However, this issue is not reflected in awareness. Figure~\ref{fig:pdrdist} shows the \ac{PDR} over the distance in the 4\,km-long Destination Area. Lines overlap for most of the distance, up to the last segment where they fan out in favor of higher penetration rates. However, this phenomenon is due to an unbalance in the turnover rate (i.e., the ratio between vehicles entering and exiting the Destination Area after the \ac{DENM} was generated). These extra vehicles are accounted for since the message is still within validity, and it is relevant to newcomers into the Destination Area.  

The main takeaway of this experiment is that, as long as Release 1 and Release 2 GeoNetworking messages are mutually intelligible, there is an effect on efficiency but not in safety (for the case of multi-hop \acp{DENM} from a single source). However, inefficient forwarding will occupy the medium with unnecessary repetitions of messages. Thus, in scenarios where there is more than one source trying to disseminate safety-critical messages, unnecessary transmissions are bound to cause collisions or, at least, to block access to the medium for more necessary messages waiting to be forwarded. Further work needs to be performed on how non-mutually intelligible messages affect performance, since Release 1 is likely to reach higher \ac{PDR} using brute force, while Release 2 will either yield access to the medium or might find a path to transmit immediately. What is sure is that, in that scenario, safety will be compromised.

\section{Discussion}
\label{sec:discussion}

We present a study of how the coexistence of two different releases of a protocol, one being an incremental improvement of the other, affects efficiency. For our case, packets were compatible, and Release 1 nodes could understand Release 2 messages and vice versa. However, the road to full \ac{CCAM} is long, and this might not be the case even in the near future. In this section, we present a discussion on the upcoming scenarios when multiple types and generations of technologies have to coexist.

\subsection{The upcoming Tower of Babel}

Vehicles equipped with \ac{ETSI} ITS nodes are already on the road communicating with large deployments. Just in the first three quarters of 2023, above 250,000 \ac{C-ITS}-equipped Volkswagen ID. cars were delivered~\cite{vwsales}. These cars can communicate with each other, with other ETSI~ITS-compatible vehicles, and with current deployments such as the one covering the entire Austrian motorway network~\cite{austriaroads}.

However, these vehicles and deployments all use Release 1 services. While some Release 2 features are software-dependant, e.g., new services such as the \ac{VA} basic service, and can be installed during car services or using over-the-air updates, some others will likely require a deeper update (e.g., compatibility with \ac{MCO}).

While backwards-compatibility is a common issue in computer networks, the characteristics of the vehicular market and industry make it specially more difficult. This is one of the first cases where a massive number of \textit{legacy} nodes will likely share spaces with nodes up to 20 years more modern~\cite{fleetage}. This will create a scenario where pockets of segregated nodes are bound to destabilize the system, at the very least make it more inefficient, while compromising efficacy and safety.

\subsubsection{Past experiences with backwards compatibility}
One example of backwards compatibility is the jump between \ac{TLS} 1.2 and 1.3. The 1.3 version was released in RFC 8446 in August 2018~\cite{rfc8446}. Its benefits over past versions have been widely studied~\cite{tlsbenefits}, but there are known examples of problems with its adoption~\cite{tlsconext}.

The main problem with \ac{TLS} 1.3 is \textit{protocol ossification}. This phenomenon occurs when deployed equipment (e.g., middleboxes) does not recognize new protocols or even extensions to known protocols that were released after they were installed. This causes them to interrupt packets that are valid, but unrecognizable for the middlebox.

The solution for \ac{TLS} 1.3, and for other examples of ossification, was to encapsulate new messages so that the \textit{wire image} of the packets is acceptable for older middleboxes. This could be a path to follow with safety-critical messages exchanged by nodes executing different releases of \ac{ETSI} ITS.

At the Access layer, 802.11p (upon which \ac{ETSI} ITS-G5 is based) and its evolution 802.11bd are somewhat compatible. One of the main differences between 802.11bd and 802.11p is the channel bandwidth --- 20\,MHz up from 11p's 10\,MHz. However, 11bd can also work in 10\,MHz, and does so if it detects nodes using only 10\,MHz, thus, falling back into 11p when needed. However, this approach might not be efficient in Future Mobility scenarios, when 11p's channel capacity might not be able to accommodate the myriad of applications that will try to use the medium. 

The foreseeable scenario if nodes cannot process packets from newer releases (i.e., if Release 1 nodes cannot handle Release 2 GeoNetworking traffic) can cause a disruption in Non-Area GeoNetworking~\cite{etsiNewGeoNetworking} if Greedy Forwarding is used. Since it is likely that beacons (e.g., CAMs) will always be compatible, a Release 2 node can select a Release 1 neighbor as the next hop for a message. The next hop will not process the message, and thus it will not reach the Destination Area, since there are no fallback nodes in ETSI Greedy Forwarding. This phenomenon can be avoided, for example, using \ac{CBF}, where multiple nodes become the next hop and contend to forward the message, increasing the chances of nodes from both releases hearing the forwarded message. Further work will address the impact of this phenomenon on Non-Area forwarding.

\subsubsection{Nodes with different technologies}

In the network side, even at Day 1, there is an identified risk of \textit{non-interoperability}~\cite{c-roads-interops}. Since \ac{ETSI} ITS is media-independent, it does not mandate that one access technology shall be used. Thus, there are vehicles and road-side equipment that use, e.g., LTE or 802.11p. \ac{ETSI} recognizes the scenario and proposes co-existance methods~\cite{etsiCoex} where, for example, vehicles using different technologies share the time domain. This means that cellular-based nodes occupy the \ac{C-ITS} band for a fraction of the time and WiFi-based nodes use it for the complement. This, however, is not full inter-operability, since nodes using different access technologies will not "listen" to each other, and this approach compromises every metric: efficiency (diminishing the amount of resources), efficacy (messages are not delivered to all connected road users), and thus, safety.

Further work has to be performed within the research and industry communities to 1) determine whether WiFi and cellular can possibly inter-operate, and 2) whether inter-operability is possible, search for a path to evolve in a way that newer versions of access technologies account for older nodes. One possible approach is to adopt approaches such as Software-defined Radio (SDR), which would allow equipment to be updated over the air as long as hardware supports newer features, such as different modulation and coding schemes.

This phenomenon will be aggravated when technologies from different Days coexist. For example, a \textit{legacy} node that cannot interpret or even receive intention-sharing or maneuver-coordination message exchanges will likely affect the way \ac{CCAM}-enabled vehicles converge to a solution. Once again, this will affect traffic efficiency and might hinder road safety. Further work is being performed to assess the effect of a mixed fleet in the optimal performance in \ac{CCAM}.

\subsection{The case for ETSI CBF Release 2}
For the specific phenomenon in this work, the differences between \ac{ETSI} \ac{CBF} Release 1 and 2 are purely software-based. There is no need for extra fields in the headers, or new values in the existing fields. The main differences come in what the algorithm does with information it already used, namely, the position vector from the last hop and the source. It also uses an existing interface to the Management entity to consult the cross-layer \ac{DCC} mechanism and account for transmission rate control information when calculating a contention timer (although this feature is optional). 

We foresee two simple solutions: 1) existing equipment that is able to receive an update adopts Release 2, or 2) Release 2 GeoNetworking messages are encapsulated as Release 1, as was the case for \ac{TLS} 1.3. Both approaches will ensure safety in given scenarios, but approach 1 guarantees more efficiency, and thus, more availability of resources for other applications.

\section{Conclusions and Future Work}
\label{sec:conclusion}

We have presented a study of the coexistence of two releases of a GeoNetworking protocol in the context of \ac{ETSI} ITS --- Release 1 and 2 of \ac{ETSI} \ac{CBF}. We have proved that, as long as releases are compatible and nodes can understand each other, safety metrics stay high even if resource efficiency is compromised. Then, we presented a discussion of possible settings that are likely to happen when Future Mobility is completely mature (i.e., Day 4 of Vision Zero), where a Tower of Babel scenario might occur, and road users are segregated into pockets of nodes \textit{speaking} different \textit{languages} (and some not \textit{speaking} at all). Even when the first C in \ac{CCAM} stands for \textit{cooperative}, this cooperation is not likely to occur when agents are not able to hear and understand each other. Future work includes a more in-depth analysis of the effect of \textit{multi-modal road users} (e.g., disconnected users, legacy fleet) in the optimal performance of the \ac{CCAM}-enabled fleet (i.e., connected and automated vehicles).

\section*{Acknowledgment}
This work was partially supported by SAFER in the project ``Human Factors, Risks and Optimal Performance in Cooperative, Connected and Automated Mobility'', the Knowledge Foundation in the project “SafeSmart – Safety of Connected Intelligent Vehicles in Smart Cities” (2019-2024), and the ELLIIT Strategic Research Network in the project ``6G wireless'' – sub-project ``vehicular communications''

\bibliographystyle{IEEEtran}
\bibliography{coexistence}

\end{document}